\documentclass[a4paper,11pt]{article}
\usepackage{graphicx}
\usepackage{color}

\addtocounter{section}{0}

\newcommand{\be}{\begin{equation}}
\newcommand{\ee}{\end{equation}}
\newcommand{\bea}{\begin{eqnarray}}
\newcommand{\eea}{\end{eqnarray}}

\definecolor{brique}{rgb}{.9,.2,0}
\definecolor{blvert}{rgb}{0,.8,.85}
\definecolor{vertcl}{rgb}{0,1,.7}

\begin{document}

\pagestyle{empty}
\setcounter{page}{0}
\hspace{-1cm}
\begin{center}
{\LARGE Network Models: Action formulation}
\end{center}
\hspace{1cm}
\begin{center}

{\bf Sh.~Khachatryan\footnote{e-mail:{\sl
shah@mail.yerphi.am}}$^{,a}$,~~  A.~Sedrakyan\footnote{e-mail:{\sl sedrak@nbi.dk}}$^{,a}$
and ~~P.~Sorba\footnote{e-mail:{\sl sorba@lapp.in2p3.fr}}$^{,b,c}$}\\

\vspace{1cm}

{\it $^a$ Yerevan Physics Institute, Alikhanian Br. str. 2, Yerevan 36,  Armenia\\

 $^b$ LAPTH, Universit\'e de Savoie, CNRS, 9 Chemin de Bellevue, BP110, 74941, Annecy-le-Vieux CEDEX, France\\

 $^c$ CERN - Theory Division, CH1211, Geneva 23, Switzerland \hspace{3cm}}

\end{center}

\vskip 0.7 truecm

\vspace{20mm}

\centerline{{\bf Abstract}}

We develop a technique to formulate quantum field theory on
 arbitrary network, based  on different, randomly disposed sets of
scattering's.  We define R-matrix of the whole network as a product
of R-matrices attached to each of scattering nods. Then an
action for a network in terms of
fermionic fields is formulated, which allows to calculate the transition
amplitudes as their Green functions. On so-called bubble and triangle diagrams
it is shown that the method
produces the same results as the one which uses the generalized
star product.
The approach allows to extend network models by including multiparticle
interactions at the scattering nods.

\vspace{5mm}

\indent

\vfill
\rightline{LAPTH-1324/09} \rightline{CERN-PH-TH/2009-041}
\newpage
\pagestyle{plain} \renewcommand{\thefootnote}{\arabic{footnote}}


\section{Introduction}
Special and growing attention is brought today to quantum graphs,
that is networks of one dimension and thin quantum wires
connected at nodes. If such an interest can easily be explained by
the current developments in nano-scale technology, one may also
note that this domain of research is particularly attractive as
well for condensed matter physicists as for mathematical graph theory
\footnote{see for example\cite{Proc} for a recent
and detailed overview on this subject}.

Different aspects of quantum field theory (QFT) have to be taken
into account and developed for such a context of problem, with among
others the behavior of fields at the junctions(vertices) of different edges.
Up to now, it is the case of star graphs which has
received special consideration and for which some significant
results have been obtained. Such graphs are constituted by only
one vertex connecting several edges, and can be seen as the
fundamental parts of generic graphs. The case of a junction with
three quantum wires has in particular been considered in the study
of a one-dimensional electron gas \cite{Nayak, Barnabe, Oshikawa, DRS}
 with the technique of bosonisation
naturally playing an essential role.

Meantime, the formulation of quantum scattering theory on a graph,
based on the connection of self adjoint extensions of the
Schrodinger operator and Hermitian symplectic forms, provided an
explicit expression of the unitary scattering S-matrix formed by
the reflection and transmission amplitudes and encoding the
allowed boundary conditions at the vertices \cite{KS:1999}.
Moreover, in the works \cite{KS:1999, KS1, KS2} different, but equivalent
rules for obtaining the transmission
and  reflection amplitudes were given. One rule
is based on the so called
generalized star product. All rules are based on
linear algebra and are relatively simple to apply. In particular
they may easily be implemented
for the purpose of numerical calculations.

 Then,
combining these last results with an algebraic field theoretical
treatment for quantum integrable systems in (1+1) dimensions with
a reflecting and transmitting impurity \cite{MRS} allowed the
authors of \cite{BM,BMS,BBMS} to develop a general
framework for constructing, via the so-called
reflection-transmission or "R-T" algebra, vertex operators from
bosonic fields propagating freely in the bulk and interacting at
the junction. After a detailed study of scale-invariant
interactions, and an explicit determination of the critical points
for a star graph with any number $n$ of edges, the four fermion
bulk interaction could then be considered and the
Tomonaga-Luttinger/Thirring model solved at such critical points.
This result is finally used for determining the charge and spin
transport as well as to establish a relationship between them,
recovering and extending the previous results first obtained
in \cite{Nayak}, and later confirmed in \cite{Barnabe, Oshikawa, DRS}.

Such interesting results urge on the study of more complex graphs
representing more realistic situations, the rather obvious next
step being the case of a network with more than one vertex, linked
together by internal wires, and particularly configurations where
closed loops are present. The simplest examples in this last case
are the "bubble diagram" (see below Fig.7) with two vertices
related by two internal wires, and the "triangle diagram" (see
Fig.8) where three star graphs combine in such a way that each
couple of vertices is connected by an internal line. In order to
extend the previous formalism developed in
\cite{BM, BMS, BBMS}, one can wonder whether the
closed diagram made of several junction points, each one connected
with an internal as well as one or more external wires, can be
represented by an unique "mega" vertex with the same external
lines. This is one of the results of this paper, where an explicit
construction of the scattering S-matrix relative to the "mega"
vertex is given in terms of the scattering S-matrix elements of
the different  vertices involved. At this point let us mention that
direct computations of the final S-matrices have recently been
performed in the case of a chain of impurities Ref.\cite{MR} as well
as for closed loops Ref.\cite{R}. A direct comparison of some of
the results obtained by this last approach with ours can easily be
done. But we must stress that the technique we are using is
completely different from the one of Ref.\cite{MR, R} and
will allow us, as explained below, to propose an action
formulation for models defined on a lattice associated to the
considered network.

It is the extension of transfer matrix approach which is hereafter used to
handle this problem. More precisely, to each of vertexes of the network
we allocate a R-matrix, which is defined by the corresponding
scattering S-matrix. Then we construct a product of all R-matrices
of the network in an appropriate way defining mega-R-matrix.
The resultant R-matrix defines an S-matrix of the whole network.
This is one of main results of the paper. Then,
using an approach first
initiated in Ref.\cite{Sed}, we rewrite the R-matrix constructed for each
star graph in terms of creation /annihilation operators
( representing the in/out-coming particles)  and with
coefficients directly related with S-matrix entries. The
replacement of such operators by Grassmann-valued fields in 2-dim.
space allows one to perform directly the product of such
R-matrices, each associated to a different vertex connected to the
next one by (at least) one internal line, and finally to
reconstruct from the so-obtained transfer matrix the one mimicking
the scattering of the "mega" vertex containing the different
connected junctions.

This technique, namely the transformation of the R-matrix into its
coherent states basis in fermionic Fock-space allows to give an
action formulation of the corresponding model on a lattice formed
by the network. The explicit check of equivalence of the action formulation of network models
on a basis of R-matrix with the star-product rules of calculations, earlier formulated
 in the paper \cite{KS1}, for so-called bubble and triangle diagrams
 is the another important  result of this article.  Note that the general proof of equivalence
is given in \cite{KSS}.

The action formulation
is a convenient way, if not the only, to investigate networks  with
large number N of vertices, or near the critical point, when one can
develop an equivalent quantum field theory approach to the problem.

The importance of the action formulation of network models is hard
to overestimate. It can be applied, for example, to the
Chalker-Coddington (CC) phenomenological model \cite{CC} devoted
to plateau-plateau transitions in quantum Hall effect. This
approach to the CC model was developed in \cite{AS}.

Let us at this point note that in Ref.\cite{Sed} such a
fermionization technique was also used for the formulation of
lattice models in (2+1) dim. in connection with the 3D-Ising
model.
 Applications for the construction of integrable models
with staggered parameters have also been found \cite{ARN}; see
also \cite{AKhS} for more developments. Obviously, the  direct
connection between  such lattice models and quantum graphs has to
be exploited  more deeply, and we will return to this point in
our conclusion.

The paper is organized as follows. We start in Section 2 by
considering the case of several impurities on a line. Then one
easily notices that, by simply re-organizing the entries of the
scattering S-matrix relative to each vertex, one gets transfer
(T-) matrices such that their product, following the order of the
different vertices along the line, provides one with a transfer
matrix relating the external legs of the diagram: it is then a
simple exercise to deduce, by an operation inverse to the previous
one, a matrix which can be seen as the S-matrix of the "mega"
vertex of the diagram. Actually,the T-matrix associated to each
vertex can be seen as the one particle sector of the extended
XX-model R-matrix.

In the Section 3 we give the formulation of three-channel
$R^{(3)}$-matrix in a standard matrix and  operator forms. Then we
pass to coherent state basis and present a simple expression of
the $R^{(3)}$-matrix  via three channel $S$-matrix. Then  we formulate a
problem of complicated networks and express  total $R$-matrix as a
product of its ingredient $R^{(2)}$ and $R^{(3)}$ matrices. This
allows us also to give an action formulation for any complicated
networks in via fermionic fields.

In Section 4 we apply our method to calculation of the so-called
bubble diagram, while in Section 5 we present the calculations for
the triangle vertex diagram and show, that corresponding mega-S-matrices
coincides with the results obtained via star-product procedure given
in  \cite{KS:1999, KS1}.

\section{Two particle R-matrix}
\subsection{Connection of Transfer matrix with R-matrix}
The act of the scattering of the particle on a potential
center/impurity can be described by  a 2x2 S-matrix of the form
\begin{eqnarray}
 \label{2-s-matrix}
\mid i >_{out} = S_{i j} \mid j>_{in},
\end{eqnarray}
where $i=1,2$ denote the number of channels: Figure 1 demonstrates
this process.
\begin{figure}[ht]
\centerline{\includegraphics[width=85mm,angle=0,clip]{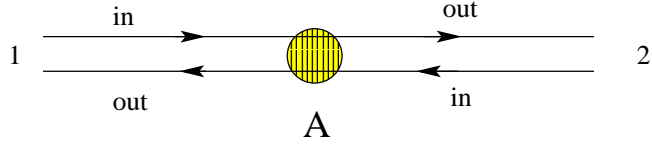}}
\caption{Two channel S-matrix.}
\end{figure}

 The S-matrix consists of transmission
$T_{12}(k),\; T_{21}(k)$ and reflection $R_{11}(k),\;
R_{22}(k)$ amplitudes ( details can be found
 in  \cite{KS:1999, MR, MSo}.
\begin{eqnarray}
 \label{2-s-matrix-2}
S=\left( \begin{array}{cc} R_{11}(k) & T_{12}(k)
\\T_{21}(k) & R_{22}(k)
\end{array}
\right).
\end{eqnarray}

This two particle process can be described also by the transfer
matrix which maps the states $\mid 1>_{in}, \mid 1>_{out}$ on the
left hand side of the scattering region $A$ to the $\mid 2>_{in},
\mid 2>_{out}$ on the right hand side.
\begin{eqnarray}
\label{2-transfer-matrix} \left( \begin{array}{c}
\mid 2>_{in}\\
\mid 2>_{out}
\end{array}\right)=
\left( \begin{array}{cc}
\frac{1}{T_{12}(k)}&-\frac{R_{11}(k)}{T_{12}(k)}\\
-\frac{R_{11}(-k)}{T_{12}(-k)}&\frac{1}{T_{12}(-k)}
\end{array}\right)
\left( \begin{array}{c}
\mid 1>_{out}\\
\mid 1>_{in}
\end{array}\right)
\end{eqnarray}

In (\ref{2-transfer-matrix})  the symmetry properties (unitarity, analyticity,
consistency) of the $S$-matrix (\ref{2-s-matrix-2})  have been taken into account.
 \cite{KS:1999, MSo,MR}.

Since the transfer matrix maps the states on left hand side to the
states on the right hand side of the scattering region the process
of scattering on the multiple points will simply be described by
the multiplication of the transfer matrices (see Fig.2).
\begin{figure}[ht]
\centerline{\includegraphics[width=125mm,angle=0,clip]{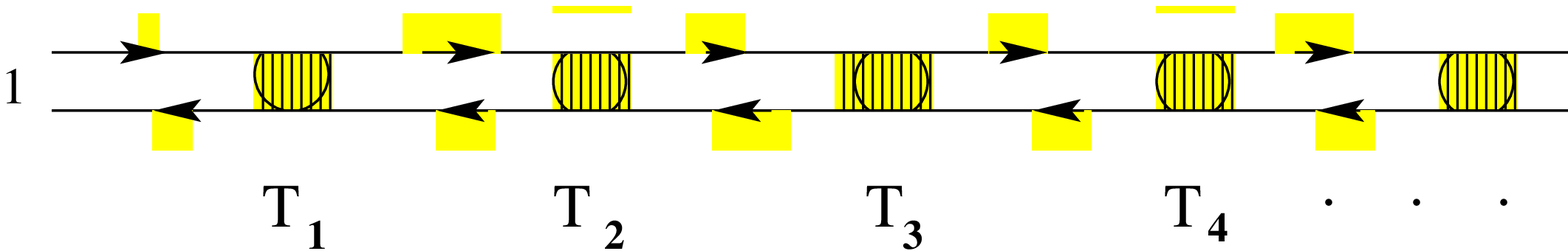}}
\caption{Chain of two channel T-matrices representing a chain of
equidistant impurities.}
\end{figure}
Continuously mapping the states from the left to the right on the
Fig.2, we will have a product of transfer matrices
\begin{eqnarray}
\label{tr-mat} T^{(n)}=T_n T_{n-1} ...T_1,
\end{eqnarray}
the matrix elements defining final transmission $T^{(n)}_{12}(k)$
and reflection $R^{(n)}_{11}(k)$ amplitudes. This procedure of
calculation of amplitudes is much simpler than calculations in
S-matrix formalism, where we do not have local matrix
multiplication but should deal with counting of non-local effects.

It appears that this transfer matrix can be considered as an one
particle sector of the R-matrix of an XX-model. Namely, the
standard form of the extended XX-model's R-matrix can be
represented as:
\begin{eqnarray}
 \label{XX}
R_{XX}=\left(\begin{array}{cccc}
-r_{22}&0&0&0\\
0&1&-r_{21}&0\\
0&r_{12}&r_{11}r_{22}-r_{12}r_{21}&0\\
0&0&0&-r_{11}
 \end{array}
\right),
\end{eqnarray}
where we denoted
\begin{eqnarray}
 \label{RXX}
R_{01}^{01}&=&1,\;\;\qquad R_{10}^{10}=r_{11}r_{22}-r_{12}r_{21},\;\;R_{11}^{11}=-r_{22}\nonumber\\
R_{00}^{00}&=&-r_{22},\;\; \qquad R_{01}^{10}=r_{12},\;\qquad
\qquad R_{10}^{01}=-r_{21}.
\end{eqnarray}
Then, by use of unitary properties of the S-matrix
(\ref{2-s-matrix-2}) one easily can see, that the middle block of
elements $R_{10}^{10},\; R_{01}^{01},\; R_{01}^{10},\;
R_{10}^{01}$ will coincide with the transfer matrix
(\ref{2-transfer-matrix}) after dividing them on $-R_{00}^{00}$
\begin{eqnarray}
\label{tr-2} T_{21}(k)&=&r_{11},\;\; T_{12}(k)=r_{22},\;\;
\nonumber\\
R_{11}(k)&=&r_{21}, \;\; \quad R_{22}(k)=r_{12}.
\end{eqnarray}
Hence, the $r$-matrix elements are connected with the $S$-matrix
ones as:
\begin{eqnarray}
 \label{tr-2}r = C \cdot S
\end{eqnarray}
with
$C=\left(\begin{array}{cc}
0&1\\
1& 0
\end{array} \right)$,
 the numeration of spaces of the initial states differing in the R and T matrix
 formulations (compare Fig.1 with Fig.3 below).

Via Jordan-Wigner transformation, the R-matrix (\ref{XX}) can be
represented as an operator consisting of fermionic
creation/annihilation operators which act on a corresponding Fock
space made of two quantum spaces (see Fig.3 below).
\begin{eqnarray}
 \label{VV}
R^{(2)}_{12}:  V_1 \otimes V_2  \rightarrow V_1 \otimes V_2.
\end{eqnarray}

Now, let us introduce scalar fermions $c^+_i,\; c_i$ in each
linear space $V_i,\;\; i=1,2$ forming a set of quantum states $| i
>_{in/out}$. Defining the R-operator as
\begin{eqnarray}
 \label{2-operator-R}
R^{(2)}_{12}= :e^{(-1)^j c_i^+ (\delta_{ij}+r_{ij})c_j} :,
\end{eqnarray}
where we have normal ordering convention of operators in space
$V_1$ and anti-normal ordering convention space $V_2$, one will
obtain \cite{Sedrakyan-11, Wadati}
\begin{eqnarray}
 \label{2d-operator-R}
R^{(2)}_{12}&=&R_{11}^{11} n_1 n_2+R_{00}^{00} \bar{n}_1 \bar{n}_2
+ R_{10}^{10} n_1 \bar{n}_2
\nonumber\\
&+& R_{01}^{01} \bar{n}_1 n_2+ R_{10}^{01} c^+_1 c_2 + R_{01}^{10}
c^+_2 c_1,
\end{eqnarray}
where $n=c^+_i c_i,\;\; i=1,2$ is the particle number operator and
$\bar n_i=1-n_i$.

\subsection{Chain of Equidistant Impurities}
The transfer matrix approach allows to analyze very easily a chain
of scattering centers/impurities and to calculate the
transmission/reflection amplitudes. In this section we will
explicit such a calculation by considering a chain of equidistant,
identical impurities.

For such a purpose, we start by defining the local transfer matrix
$T_i$ at the chain point $i$, with $i = 1,...,n$ by translating
the transfer matrix $T_0$ defined by the formula
(\ref{2-transfer-matrix}) at the point zero via translation
operator
\begin{eqnarray}
 \label{translation}
W_i=\left(\begin{array}{cc}
e^{i k x_i}&0\\
0& e^{-i k x_i}
\end{array} \right)
\end{eqnarray}
as  $ T_i = W_i^{-1} T_0  W_i$. Then the final transfer matrix
appears as the product of the different local transfer matrices
$T_i$:
\begin{eqnarray}
\label{chain-trm} T^{(n)}= T_n T_{n-1}  \dots T_1 = W_{n}^{-1}( T_0 W)^n W^{-1} W_1,
\end{eqnarray}
where $W= W_{i+1} W_{i}^{-1}$ is constant due to equidistant
position of impurities $d=x_{i+1}-x_i=const$.

We need to calculate $T^{(n)}$. In order to reach this goal one
can diagonalize the matrix $T_0 W$ with the help of an unitary
matrix $U$:
\begin{eqnarray}
\label{n-degree}
 T^{(n)}=W_{n}^{-1}U ( T_0 W)^n_{diagonal}U^+ W^{-1}W_{1}
\end{eqnarray}

The eigenvalues and eigenvectors of $T_0 W$ can be obtained
via:
\begin{eqnarray}
\label{U} \left( \begin{array}{cc}
\frac{e^{i k d}}{T_{12}(k)}&-\frac{R_{11}(k) e^{-i k d}}{T_{12}(k)}\\
-\frac{R_{11}(-k) e^{i k d}}{T_{12}(-k)}&\frac{e^{-i k d}}{T_{12}(-k)}
\end{array}\right)
\left(\begin{array}{c}
u_{\pm}\\
v_{\pm}
\end{array}\right)=\lambda_{\pm}\left(\begin{array}{c}
u_{\pm}\\
v_{\pm}
\end{array}\right),
\end{eqnarray}
where  eigenvalues  $\lambda_{\pm}$ easily can be found with the
characteristic equation:
\begin{eqnarray}
\label{eigen-0} \lambda^2 -\lambda \left( \frac{e^{i k
d}}{T_{12}(k)}+\frac{e^{-i k d}}{T_{12}(-k)}\right)+1=0.
\end{eqnarray}

Defining
\begin{eqnarray}
\label{eigen-12} \alpha(k)= ArcCosh\left[\frac{1}{2}\left(
\frac{e^{i k d}}{T_{12}(k)}+\frac{e^{-i k
d}}{T_{12}(-k)}\right)\right]
\end{eqnarray}
one can represent eigenvalues in a form
\begin{eqnarray}
 \label{eigen-13}
\lambda_{\pm}=e^{\pm \alpha(k)}.
\end{eqnarray}

After some more algebra, one gets for $T^{(n)}$:


\begin{eqnarray}
 \label{transfern}
T^{(n)}= \left(
\begin{array}{cc}
 \frac{\sinh(n\alpha)}{\sinh(\alpha)} \frac{1}{T_{12}(k)}e^{i
k(n-1)d}
 &  -
\frac{\sinh(n\alpha)}{\sinh(\alpha)}
\frac{R_{11}(k)}{T_{12}(k)}e^{-i k(2 x_1+(n-1)d)}
\\
- \frac{\sinh((n-1)\alpha)}{\sinh(\alpha)}e^{i k(n-2)d} & \\[5mm]
- \frac{\sinh(n\alpha)}{\sinh(\alpha)}
\frac{R_{11}(-k)}{T_{12}(-k)}e^{i k(2 x_1+(n-1)d)} &
\frac{\sinh(n\alpha)}{\sinh(\alpha)} \frac{1}{T_{12}(-k)}e^{-i
k(n-1)d}
\\
& - \frac{\sinh((n-1)\alpha)}{\sinh(\alpha)}e^{-i k(n-2)d}
\end{array}
\right).
\end{eqnarray}

Note that, in the case of only one impurity, i.e. n=1 and d=0, one
directly recovers the T-matrix defined in Eq. (2.3) - as it should
be.

\subsection{Fermionic Coherent States}

In this subsection we will define fermionic coherent states which
are very instrumental in constructing of action/Lagrangian
formulation of network models \cite{Sed}. They form another basis
of the linear space of the quantum states, giving the possibility
to express the matrix elements of the R-matrix via Grassmann
variables.

By definition,  fermionic coherent/anticoherent states are
eigenstates of fermion creation/annihilation operators $c, c^+$.
Since the operators $c, c^+$ have anti-commuting properties, their
eigenvalues cannot be ordinary numbers. They will be Grassmann
variables satisfying the following properties:
\begin{eqnarray}
 \label{grassmann}
\psi \psi&=&\bar\psi \bar\psi=0, \qquad \qquad \psi \bar\psi= - \bar\psi \psi,\nonumber\\
\int d\psi&=& \int d\bar\psi=0, \qquad \int d\psi \psi =\int
d\bar\psi \bar\psi=1.
\end{eqnarray}
With the help of these variables, let us define a coherent state:
\begin{eqnarray}
\label{coh-s1} \mid \psi\rangle = e^{- \psi c^+}\mid 0 \rangle,
\qquad \qquad \langle \bar{\psi}\mid= \langle 0\mid e^{-c
\bar{\psi}},
\end{eqnarray}
as being an eigenstate of the annihilation operator $c$ with the
eigenvalue $\psi$:
\begin{eqnarray}
\label{eigen-1} c\mid \psi\rangle =\psi\mid \psi\rangle
  \qquad \qquad
\langle\bar{\psi}\mid c^+=\langle\bar{\psi}\mid\bar{\psi}.
\end{eqnarray}
Correspondingly, the anti-coherent state
\begin{eqnarray}
\label{coh-s2} \mid \bar\psi\rangle = (c^+-\bar\psi)\mid 0
\rangle,  \qquad \qquad \langle \psi\mid= \langle 0\mid (c+\psi),
\end{eqnarray}
is an eigenstate of the creation operator $c^+$ with the
eigenvalue $\bar\psi$:
\begin{eqnarray}
\label{eigen-2} c^+\mid \bar\psi\rangle =\bar\psi\mid
\bar\psi\rangle   \qquad \qquad \quad \; \langle\psi\mid
c=-\langle\psi\mid \psi.
\end{eqnarray}
By use of the relations (\ref{grassmann}), it is straightforward
to check that we get the following normalization:
\begin{eqnarray}
\label{normal} \langle \bar\psi\mid \psi\rangle =e^{\bar\psi \psi}
\qquad \qquad \langle\psi\mid \bar\psi\rangle =e^{\bar\psi \psi}
\end{eqnarray}
and the completeness relations:
\begin{eqnarray}
 \label{comp}
\int d\bar\psi d\psi \mid \psi\rangle\langle\bar\psi \mid e^{\psi
\bar\psi}=1, \qquad \int d\bar\psi d\psi \mid
\bar\psi\rangle\langle\psi \mid e^{\psi \bar\psi}=1
\end{eqnarray}
for the coherent and anti-coherent states respectively.

The coherent states are forming a new basis in the two dimensional
Fock space of fermions, where it becomes very easy to calculate
the matrix elements of any functional in the operators $c, c^+$
written in a normal/anti-normal ordered form. In particular, it is
straightforward to find out that, for any normal ordered operator
$:F(c^+, c):$
\begin{eqnarray}
\label{n-order1} \langle \bar\psi\mid :F(c^+, c):\mid \psi
\rangle= F(\bar\psi, \psi)e^{\bar\psi \psi}
\end{eqnarray}
while for an anti-normal ordered operator:
\begin{eqnarray}
\label{n-order2} \langle \psi\mid :F(c^+, c):\mid \bar\psi
\rangle= F(\bar\psi, -\psi)e^{\bar\psi \psi}.
\end{eqnarray}

Note also that the kernel $F(\bar\psi, \psi)$ of two operators
$:F(c^+, c):=:F_1(c^+, c)::F_2(c^+, c):$ reads:
\begin{eqnarray}
F(\bar\psi, \psi)=\int d \bar{\chi} d \chi
e^{-\bar{\chi}\chi}F_1(\bar\psi,\chi)F_2(\bar{\chi},
\psi).\label{product}
\end{eqnarray}
We will use this relation extensively below.

The R-operator (\ref{2d-operator-R}) has a very simple form in the
coherent state basis. Introducing coherent/anti-coherent states
$\bar{\psi}^{\prime}_1, \;\psi_1, \; \bar{\psi}_2, \;
\psi^{\prime}_2$ in the spaces $V_1 / V_2$ correspondingly defined
by formulas (\ref{coh-s1})/(\ref{coh-s2}), one obtains for the
matrix elements $ \langle
\psi^{\prime}_2,\bar{\psi}^{\prime}_1\mid R\mid \;\psi_1,
\bar{\psi}_2 \rangle $of the R-operator:
\begin{eqnarray}
\langle \psi^{\prime}_2,\bar{\psi}^{\prime}_1\mid R\mid \;\psi_1, \bar{\psi}_2
\rangle& =&
e^{-r_{11}\bar{\psi}_1^{\prime}\psi_1-r_{22}\bar{\psi}_2\psi_2^{\prime}
-r_{12}\bar{\psi}_1^{\prime}\psi_2^{\prime}-r_{21}\bar{\psi}_2\psi_1
\label{2d-R-coherent}
 }\nonumber\\
&=& e^{-\bar\psi_i (C S)_{ij}\psi_j}.
\end{eqnarray}
%

In order to make a correspondence with the notations on Fig.1,
where the left hand states are marked with index 1, while right
hand states are marked by 2, we redefine $\psi_1=\eta_1,
\bar{\psi'}_1=\bar{\eta}_2, \psi'_2=\eta'_2,
\bar{\psi}_2=\bar{\eta'}_1$ and obtain for the $R$-matrix the
expression
\begin{eqnarray} \label{2d-S-coherent} \langle
\eta'_2, \bar{\eta}_2\mid R\mid \;\eta_1, \bar{\eta'}_1 \rangle&
=& e^{-\bar\eta_i S_{ij}\eta_j}.
\end{eqnarray}
 Hereafter, we shall use $R$-matrices in this last formulation
like (\ref{2d-S-coherent}).

Graphically, the R-operator in the coherent state basis can be
represented as in Fig.3, where it becomes clear that the S-matrix
parameters are merely hopping parameters.

Note that, in getting formula (\ref{2d-R-coherent}) we have used
formulas (\ref{n-order1}-\ref{n-order2}) and (\ref{2-operator-R}).

\begin{figure}[ht]
\centerline{\includegraphics[width=95mm,angle=0,clip]{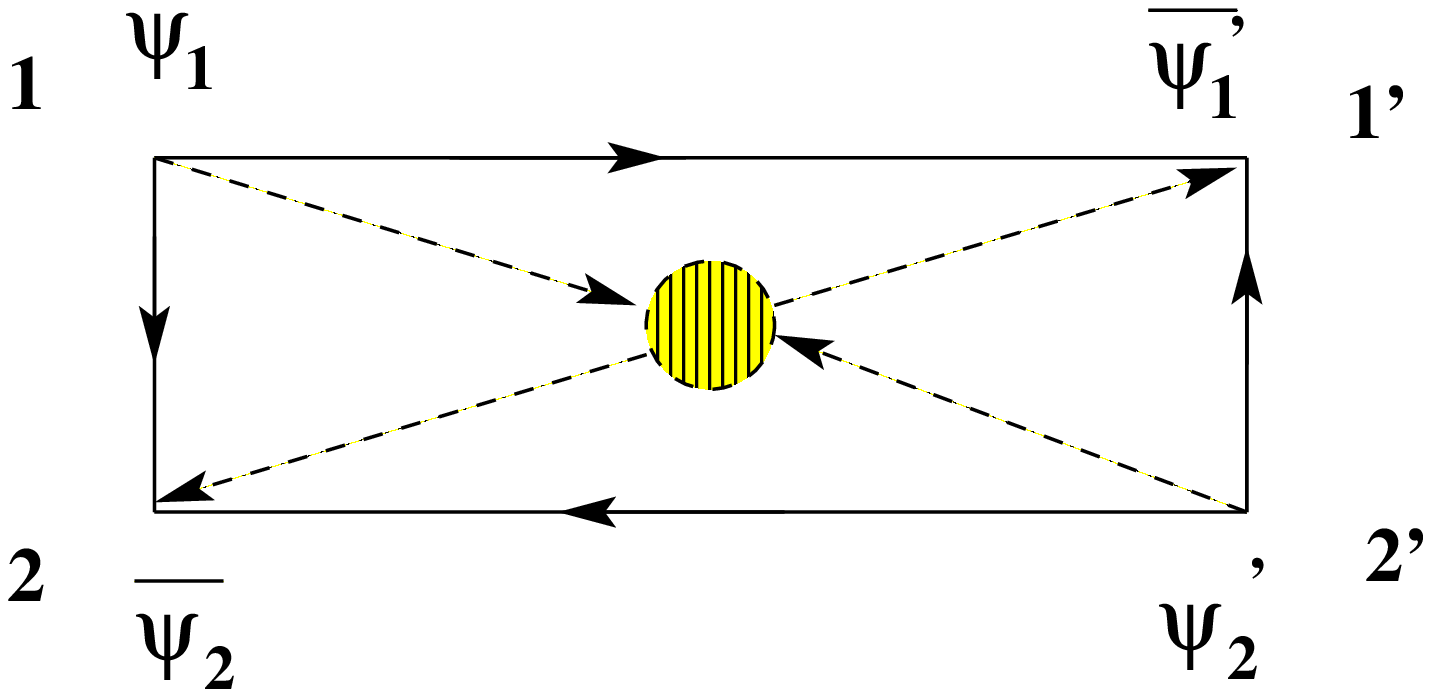}}
\caption{Convenient representation of two-channel R-matrix as an quadrangle where
the sides are hopping links of fermions.}
\end{figure}


\section{Three particle R-matrix}
\subsection{Three particle S-matrix}

In this Section we will define the three particle R-matrix
connected to the corresponding three particle S-matrix. The main
problem that we would like to solve is the following. Usually, in
order to describe the scattering of three particles we define a
three particle S-matrix. But then it is very hard to make
calculations in a situation when of a network with multiple
scattering points. The problem stands in non-local rules of the
S-matrix multiplications. The situation is totally different if we
have to deal with multiplication of R-matrices. Then we have local
multiplication rules of the tensorial type allowing to handle
complicated problems with networks.

We will start from the three particle of S-matrix and define a
corresponding R-matrix in the next section.

So, let us consider three channels where the incoming and outgoing
states are marked as $\mid i >_{in}$ and $\mid i
>_{out}, \;\; (i=1,2,3)$ respectively. In a way similar to the two particle case
(\ref{2-s-matrix}), the three particle S-matrix is an matrix
mapping incoming states to outgoing ones:
\begin{eqnarray}
 \label{S-matrix}
\mid i >_{out} = S_{i j} \mid j>_{in}
\end{eqnarray}

It is convenient to represent this scattering process by the
vertex shown in Fig.3.

\begin{figure}[h]
\centerline{\includegraphics[width=85mm,angle=0,clip]{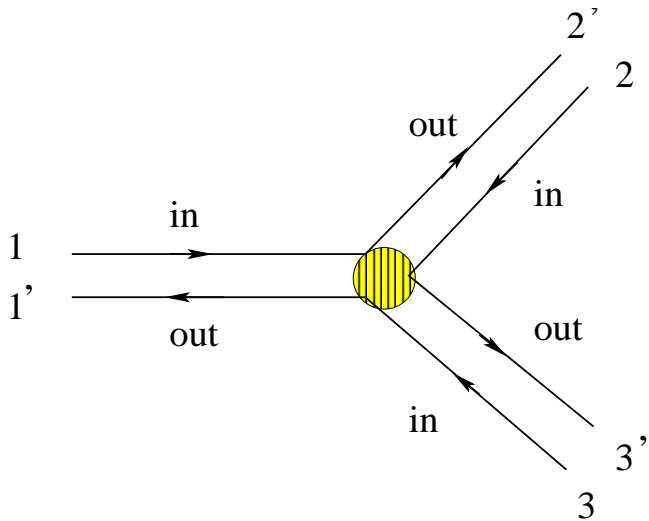}}
\caption{Three channel S-matrix/R-matrix.}
\end{figure}

\subsection{Definition of Three Particle R-matrix}

Up to this point the S-matrix was describing two channel
scattering processes where 0,1,2 particles could participate.
 Now we introduce a three
channel R-matrix containing  information about 0,1,2,3 particle
processes involved in the scattering.

For further convenience, we change the traditional notations of
the states in the channel presented in the Fig.3 and make the
re-numeration in accordance with Fig.4. Namely, let us change $|
2'\rangle \rightarrow |1'\rangle $  and $| 1'\rangle \rightarrow
|2'\rangle $ and introduce the matrix:
\begin{eqnarray}
\label{r-matrix}  r=C S,
\end{eqnarray}
where $S$ is the three channel  S-matrix (\ref{S-matrix}) and the
matrix $C$
\begin{eqnarray}\label{C-matrix}
C= \left(
     \begin{array}{ccc}
       0 & 1 & 0 \\
       1 & 0 & 0 \\
       0 & 0 & 1 \\
     \end{array}
   \right).
\end{eqnarray}
ensures the re-numeration of the states according to Fig.4 and
Fig.5.


The three particle R-matrix is a matrix/operator acting on the
direct product of the three two dimensional linear spaces $V_j,
\;\; j=1,2,3$ ( Fig.5)
\begin{equation}\label{R-matrix}
R_{123}:  V_1 \otimes V_2 \otimes V_3 \rightarrow V_1 \otimes V_2
\otimes V_3.
\end{equation}
The dimension of $V$ spaces is defined by the number of incoming
and outgoing states. In the simplest case it is two and we will
consider this case in this article. {
{Let us also define the basis states for the each space $V_i$ as
$|0\rangle_i$ and $|1\rangle_i$.}}

Our aim is the definition of the R-matrix corresponding to the
given S-matrix.
Any R-matrix can be represented in three different bases of the
linear spaces $V_i,\;i=1,2,3$. The first, and the most popular one
is an ordinary matrix form. In case of three particle, the
R-matrix entries read $R_{\alpha \beta \gamma}^{\alpha' \beta'
\gamma'}$ where each index takes the values 0 and 1. As we will
see below this notation is justified because it is connected with
the particle (fermion) number in operator representation of the
R-matrix. We will define and consider R-matrices corresponding to
the processes with particle number conservation. The most general
form of this type of R-matrix was defined in \cite{AKhS} and have
a following form:
\begin{eqnarray}
 \label{matrix-form}
R=\left( \begin{array}{cccccccc}
R_{000}^{000}&&&&&&&\\
&R_{001}^{001}&R_{010}^{001}&R_{100}^{001}&&&&\\
&R_{001}^{010}&R_{010}^{010}&R_{100}^{010}&&&&\\
&R_{001}^{100}&R_{010}^{100}&R_{100}^{100}&&&&\\
&&&&R_{011}^{011}&R_{101}^{011}&R_{110}^{011}&\\
&&&&R_{011}^{101}&R_{101}^{101}&R_{110}^{101}&\\
&&&&R_{011}^{110}&R_{101}^{110}&R_{101}^{110}&\\
&&&&&&&R_{111}^{111}
  \end{array}
 \right).
\end{eqnarray}
It is presented in a block-diagonal form, where each block
corresponds respectively to the $0-,\; 1-,\;2-,\;3-$ particle
scattering: we interpret $|0\rangle,\; |1\rangle$ states as vacuum
state and one particle state.

Each two dimensional space $V_i, i=1,2,3$ can be realized as a
Fock space of scalar fermions. This will allow to formulate the
operator R-matrix in a fermionic space equivalent to the
Jordan-Wigner transformation.

Let us introduce the creation/annihilation operators $c_i , c^+_i$
of scalar fermions in each linear space $V_i,  \; i=1,2,3$ such
as:
$$
c_i |0\rangle_i=0,\quad c^+_i|0\rangle_i=|1\rangle_i,
$$
and consider the operator \cite{AKhS}:
\begin{eqnarray}
 \label{operator-R}
R_{123}= :e^{(-1)^jc_i^+ (\delta_{ij}+r_{ij})c_j} :
\end{eqnarray}
where the matrix elements $r_{ij}$ are forming the set of
parameters of the model and are connected with S-matrix entries
via formula (\ref{r-matrix}) while the symbol :: means normal
ordering of the operators in the space $i=1,3$ and anti-normal
(hole) ordering for the fermions in $i=2$.

 Expanding the exponent in the formula (\ref{operator-R}) and
after some algebra one can obtain the following expression for the
operator R-matrix:

\begin{eqnarray}
\label{operator-R2} R_{123}&=&
R_{000}^{000}\bar{n}_1\bar{n}_2\bar{n}_3+R_{001}^{001}\bar{n}_1\bar{n}_2
n_3+
R_{010}^{010}\bar{n}_1n_2\bar{n}_3+R_{100}^{100}n_1\bar{n}_2\bar{n}_3 \nonumber\\
&+& R_{011}^{011}\bar{n}_1 n_2 n_3 + R_{101}^{101} n_1 \bar{n}_2
n_3
+R_{110}^{110} n_1 n_2\bar{n_3}+R_{111}^{111}n_1n_2n_3\nonumber\\
&+&(R_{001}^{010} c_3^+c_2+R_{001}^{010}c_2^+ c_3)\bar{n}_1
+(R_{101}^{110} c_3^+c_2+R_{110}^{101}c_2^+ c_3)n_1\nonumber\\
&+&(R_{001}^{100} c_3^+c_1+R_{100}^{001}c_1^+ c_3)\bar{n}_2
+(R_{011}^{110} c_3^+c_1+R_{110}^{011}c_1^+ c_3)n_2\nonumber\\
&+&(R_{010}^{100} c_2^+c_1+R_{100}^{010}c_1^+ c_2)\bar{n}_3
+(R_{011}^{101} c_2^+c_1+R_{101}^{011}c_1^+ c_2)n_3
\end{eqnarray}
with
\begin{eqnarray}
\label{r-matrix-1} R_{111}^{111}&=&r_{11}r_{33}-r_{13}r_{31},\;
R_{100}^{100}=r_{11}r_{22}-r_{12}r_{21},\;
R_{001}^{001}=r_{33}r_{22}-r_{23}r_{32}, \nonumber\\
 R_{010}^{010}&=&1,\;\; R_{101}^{101}=-Det[(-1)^j r_{ij}],\;\; R_{110}^{110}=-r_{11},\;R_{000}^{000}=-r_{22},\;
R_{011}^{011}=-r_{33},\nonumber\\
\nonumber\\
R_{110}^{101}&=&r_{11}r_{23}-r_{13}r_{21},\;
R_{101}^{110}=r_{12}r_{31}-r_{11}r_{32},\;
R_{101}^{011}=r_{13}r_{32}-r_{33}r_{12}, \nonumber\\
R_{011}^{101}&=&r_{33}r_{21}-r_{23}r_{31},\;
R_{100}^{001}=r_{22}r_{13}-r_{12}r_{23},\;
R_{001}^{100}=r_{22}r_{31}-r_{21}r_{32}, \nonumber\\
R_{110}^{011}&=&r_{13},\;\; \hspace{1.5 cm} R_{011}^{110}=r_{31},
\;\; \hspace{1.5 cm} R_{010}^{001}=-r_{23},\qquad \qquad\nonumber\\
R_{001}^{010}&=&r_{32},\;\; \hspace{1.5 cm} R_{100}^{010}=r_{12},
\;\; \hspace{1.5 cm} R_{010}^{100}=-r_{21}.
\end{eqnarray}

In the spaces $V_i,  \; i=1,2,3$ the creation-annihilation
operators can be identified with
 $|\alpha\rangle_i\langle\alpha'|$ as:
\bea
c_i=|0\rangle_i\langle1|,\;\;\;\;c_i^+=|1\rangle_i\langle0|,\;\;\;\;
c_i^+ c_i=|1\rangle_i\langle1|,\;\;\; c_i
c_i^+=|0\rangle_i\langle0|. \label{bra-ket}
 \eea
Note that with this definition, each $V_i$ space acquires a
grading: $|\alpha\rangle_i |\beta\rangle_j=(-1)^{p_\alpha
p_\beta}|\beta\rangle_j|\alpha\rangle_i,\;\; i\neq j$, where
$p_\alpha=\alpha$ is the parity of the state $|\alpha\rangle$.

\begin{figure}[h]
\centerline{\includegraphics[width=85mm,angle=0,clip]{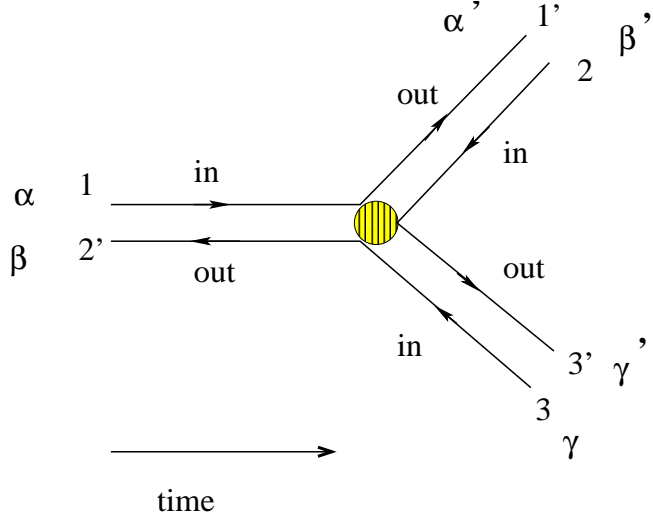}}
\caption{The R-matrix, attached to the S-matrix and defined by
this picture, has the  entries $R_{\alpha \beta\gamma}^{\alpha'
\beta' \gamma'}$}
\end{figure}

In this ket-bra language, the $R_{123}$-operator
(\ref{operator-R}) can be represented via its matrix  elements
$R_{\alpha \beta \gamma}^{\alpha' \beta' \gamma'}= \langle \gamma
\beta \alpha \mid R \mid \alpha' \beta' \gamma' \rangle $  in the
following way:
\bea \label{tildeR} R_{123}&=&R_{\alpha \beta \gamma}^{\alpha'
\beta'\gamma'}\Big(|\alpha\rangle_1|\beta\rangle_2|\gamma\rangle_3\Big)
\Big({ }_3\langle\gamma'|{
}_2\langle\beta'|{}_1\langle\alpha'|\Big)\\ \nonumber&=&R_{\alpha
\beta\gamma}^{\alpha' \beta'
\gamma'}(-1)^{p_\alpha(p_\alpha+p_{\alpha'})+p_{\beta'}(p_\gamma
+p_{\gamma'})}\Big(|\alpha\rangle_1\langle\alpha'|\Big)\Big(|\beta\rangle_2\langle\beta'|\Big)\Big(|\gamma\rangle_3\langle\gamma'|\Big).
%
\eea
The operator $R_{123}$ has even parity, and for its non-zero
matrix elements the relation
$(-1)^{p_\alpha+p_\beta+p_\gamma+p_{\alpha'}+p_{\beta'}+p_{\gamma'}}=1$
takes place.


The R-operator appears to have a very simple form in a coherent
space representation defined by formulas
(\ref{coh-s1}-\ref{coh-s2}). By using a coherent state basis
(\ref{coh-s1}) for the spaces $V_1, V_3$, and an anti-coherent
basis (\ref{coh-s2}) for $V_2$ as well as the relations
(\ref{n-order1}, \ref{n-order2}), one can obtain a very
transparent connection between the R- and the S-matrix. Namely:
\begin{eqnarray}
\label{R-coherent} R= e^{-\bar{\psi}_i
r_{ij}\psi_j}=e^{-\bar{\eta}_i S_{ij}\eta_j},
\end{eqnarray}
where we have summation over $i,j=1,2,3$ and
$\eta_i=\psi_i,\;\bar{\eta}_i= C_{ij}\bar{\psi}_j$. According to
the definitions in (\ref{r-matrix}), the first equation in
(\ref{R-coherent}) corresponds to the Fig.5, the second one to the
Fig.4, i.e $r_{ij}$  and $S_{ij}$ are the hopping parameters
defined on the network of Fig.5  and Fig.4 respectively.

{
One needs also to define the $R_{123}$ operator with inverse
orderings: hole-ordering and anti-coherent basis for the spaces
$V_1, V_3$, and normal ordering and coherent state basis for
$V_2$. This will bring to the respective changes of the
expressions of the matrix elements of $R^{(3)}$ written in terms
of $r_{ij}$ matrix elements.}

The connection of three particle $R^{(3)}$-matrix (\ref{r-matrix})
with two particle one defined by (\ref{RXX}) is very simple.
Namely:
\begin{eqnarray}
 \label{2-3}
 R_{\alpha\;\beta}^{(2) \alpha\prime\beta\prime}= R_{\alpha\;\beta\;\; 0}^{(3) \alpha\prime\beta\prime\; 0}
\end{eqnarray}

With this three-channel R-matrix $R^{(3)}$ and the two
channel R-matrix $R^{(2)}$ defined in Section 2 at hand, we are
in position to consider complicated networks with any number of
$R^{(2)}$ and $R^{(3)}$ vertices. The generalization to
multivertex ($n > 3$) processes is also straightforward: {
{ \begin{equation} R_{12...n}=:e^{\sum_{i,j}^n(-1)^{j'}c_i^+
(\delta_{ij}+r_{ij})c_j}:=\hspace{5cm}\label{rn}\end{equation}
$$=\sum_{\alpha_i, \alpha'_i}R_{\alpha_1 \alpha_2 ...\alpha_n}^{\alpha'_1 \alpha'... \alpha'_n}
\Big(|\alpha_1\rangle_1|\alpha_2\rangle_2 \cdots|\alpha_n
\rangle_n\Big) \Big({ }_n\langle\alpha'_n|\cdots{
}_2\langle\alpha'_2|{ }_1\langle\alpha'_1|\Big)$$
Here also for further convenience one can define two ways of the
orderings: normal ordering and hole ordering  for the operator
expressions respectively in the spaces $V_{2k+1}$ and $V_{2k}$ and
the inverse situation. For the first case one must take $j'=j$ in
(\ref{rn}), for the second case $j'=j+1$.}}

\subsection{Networks}

\begin{figure}[ht]
\centerline{\includegraphics[width=95mm,angle=0,clip]{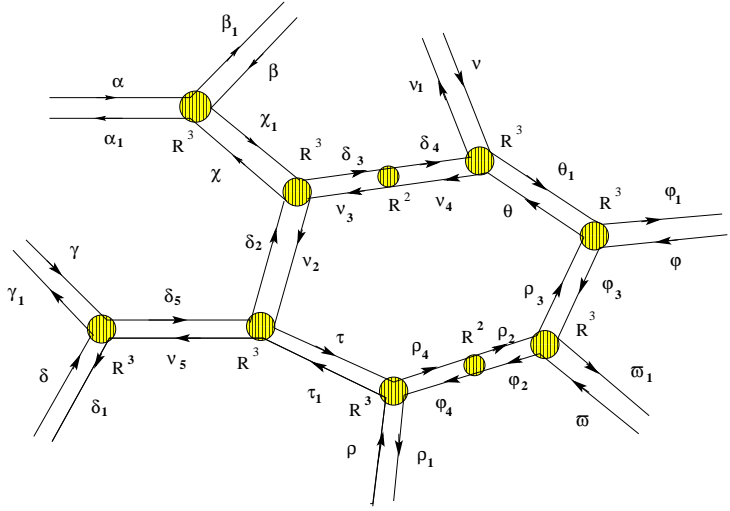}}
\caption{Example of network.} \label{net-example}
\end{figure}

In this subsection we will describe a procedure for handling a
process with multiple external channels and many $R^{(2)}$ and
$R^{(3)}$ vertices.  Our aim is to express the $R$-matrix of the
whole process via its ingredients. Let us consider now an example
of complicated network as presented  on Fig. \ref{net-example}. We
have attached the indexes $\alpha, \; \beta, \; \gamma, \;
 \delta \; \nu, \; \rho,\; \phi,\; \omega $ to the 8 incoming external states and
$\alpha_1, \; \beta_1, \; \gamma_1, \; \delta_1 \; \nu_1, \;
\rho_1\; \phi_1,\; \omega_1 $  to the 8 outgoing external states
of the network, while the internal indexes are denoted by $\chi,\;
\chi_1,\; \delta_2,\; \delta_3,\; \delta_4,\;
\delta_5,\; \nu_2,\; \nu_3,\; \nu_4,\; \nu_5, \; \theta,\;\\
\theta_1, \; \sigma,\; \sigma_1, \; \rho_2, \; \rho_3, \;
\rho_4,\; \phi_2,\; \phi_3,\; \phi_4,\;\tau,\;\tau_1$. There are
eight three particle $R^{(3)}$ and two particle $R^{(2)}$ matrices
here disposed at the points with coordinates $\xi_i,\;i=1\cdots
10$ in the space.

But the definition of the two- and three channel $R$-matrices
given above doesn't depend on the coordinates of the space; they
are defined at the zero point. Therefore,
 in order to form a network with
different positions of the scattering nodes, we should translate
each of the scattering matrices  from the zero point to their
position at $\vec{\xi}$. This can be done by use of the set of
translational matrices $W_{\vec{\xi}}(\vec{k_i})$, one for each of
channels $(i=1,2,3)$, defined as \bea \label{trans-1}
W_{\vec\xi}(\vec{k_i})= \left(\begin{array}{cc}
1& 0\\
0& e^{i \vec{k_i}\vec\xi}\end{array}\right),\;i=1,3,\quad
W_{\vec\xi}(\vec{k_2})= \left(\begin{array}{cc}
e^{i\vec{k_2}\vec\xi} & 0\\
0&1
\end{array}
\right), \eea with $\vec{k_i}= k \vec{n_i}$, where $\vec{n_i}$ is
unit vector exiting from the vertex $\vec\xi$ along a channel
$i=1,2,3$ in direction to the neighbor vertex (defined by arrows
on the figures). The matrices $W_{\vec\xi}(\vec{k_i})$ and
$W_{\vec\xi}(\vec{k_i})^{-1}=W_{\vec\xi}(-\vec{k_i})$ are acting
on the incoming and outgoing states of the channels respectively.
On the matrix elements  (\ref{r-matrix-1}) of  the $R$-matrix
 they act as:
\bea
\label{trans} R_{\alpha \beta \gamma}^{\alpha' \beta'
\gamma'}(\vec\xi) &=&(W_{\vec\xi}(\vec{k_1})^{-1})_{ \alpha}^{\alpha_1}
(W_{\vec\xi}(\vec{k_1}))_{\beta}^{\beta_1}(W_{\vec\xi}(\vec{k_3})^{-1})_{\gamma}^{\gamma_1}
R_{\alpha_1\beta_1\gamma_1}^{\alpha_2\beta_2\gamma_2}  \nonumber\\
&\cdot&
(W_{\vec\xi}(\vec{k_2}))_{\alpha_2}^{\alpha'}(W_{\vec\xi}^{-1})(\vec{k_2}))_{
\beta_2}^{\beta'}(W_{\vec\xi}(\vec{k_3})))_{\gamma_2}^{\gamma'}\quad
\eea In the fermionic operator formulation (\ref{operator-R}),
this transformation reads as
\bea \label{trans-2} R_{123}(\vec\xi) = (:e^{-(1-e^{-\vec\xi
\vec{k_1}})c^+_1 c_1}:)\; (:e^{(1-e^{\vec\xi \vec{k_1}})c^+_2
c_2}:)\;(:e^{-(1-e^{-\vec\xi \vec{k_3}})c^+_3 c_3}:)\;
 R_{123}  \nonumber\\
\times (:e^{-(1-e^{\vec\xi \vec{k_2}})c^+_1 c_1}:)\;
(:e^{(1-e^{-\vec\xi \vec{k_2}})c^+_2 c_2}:) \; (:e^{-(1-e^{\vec\xi
\vec{k_3}})c^+_3 c_3}:),\qquad \eea
 which is equivalent to the corresponding
change of the parameters $r_{ij}$ in (\ref{r-matrix-1}). Namely,
 \bea \label{r12}
r_{ij}(\vec{\xi})= e^{-(\vec{k}_i-\vec{k}_j)\vec{\xi}}
r_{ij}, \eea
where the $r_{ij}$ are defined in (\ref{r-matrix}),
(\ref{R-coherent}).

 The equation (\ref{r12}) gives also the  transformation formula
for the two-channel matrix elements of two-channel
${R^{(2)}}_{\alpha\beta}^{\alpha'\beta'}$.


With the notations of $R^{(3)}$-matrix elements
(\ref{r-matrix-1}), the resulting ${R^{(8)}}$-matrix elements,
describing the process of Fig.6,  can be written as
\begin{eqnarray}
\label{example}\nonumber &&{R^{(8)}}_{\gamma\; \gamma_1\; \delta\;
\varpi_1\; \beta\; \beta_1 \;\varphi
\;\varphi_1}^{\alpha_1\; \alpha\;\delta_1 \;\varpi\; \nu_1 \;\nu \; \rho_1\; \rho}=\frac{1}{N}\sum(-1)^{p_{\tau_1}+p_{\chi}+p_{\theta}}\\
&&{R^{(3)}}_{\gamma \gamma_1 \delta}^{\delta_5 \nu_5 \delta_1}
{R^{(3)}}_{\delta_5 \nu_5 \tau_1}^{\delta_2 \nu_2 \tau}
{R^{(3)}}_{\chi \delta_2 \nu_2 }^{ \chi_1 \delta_3 \nu_3}
{R^{(3)}}_{\chi_1 \beta \beta_1}^{ \chi \alpha_1 \alpha}
{R^{(2)}}_{\delta_3 \nu_3}^{\delta_4 \nu_4 }\\\nonumber
&&{R^{(3)}}_{\delta_4\nu_4 \theta}^{\nu_1 \nu \theta_1}
{R^{(3)}}_{ \varphi \varphi_1 \theta_1}^{\varphi_3 \rho_3 \theta }
{R^{(3)}}_{\varpi_1  \varphi_3 \rho_3}^{\varpi \varphi_2 \rho_2 }
{R^{(2)}}_{\varphi_2 \rho_2}^{\varphi_4 \rho_4}
{R^{(3)}}_{\varphi_4\rho_4 \tau }^{\rho_1\rho\tau_1}
\end{eqnarray}
 where we have summation over all
internal repeating indexes.

The normalization factor $N$ in the expression (\ref{example}) is
essential to factor out vacuum(closed loop) processes in the
network, which are going on without inclusion of external states.
This factor will be concretely specified in examples presented in
the next two sections. The signs $(-1)^{p}$ are reflecting the
graded character of the $R$-operators (\ref{rn}): for recovering
the formula (\ref{example}), one has to make the product of the
 operators (\ref{rn}) corresponding to each vertex and to sum up over
 all the internal states.

 One must also takes into account the
following multiplication rules: if
the product of two operators relative to a given internal channel
$n$ contains a product such as ${ }_n\langle\alpha|\times
|\bar{\alpha}\rangle_n$, it must be evaluated using ${
}_n\langle\alpha|\bar{\alpha}\rangle_n=\delta_{\alpha\bar{\alpha}}$,
and if a product like $|\beta\rangle_n\times {
}_n\langle\bar{\beta}|$ occurs, then it can be derived using:
$|\beta\rangle_n\langle\bar{\beta}|=(-1)^{p_{\beta}}\delta_{\beta\bar{\beta}}$.
This second relation can be obtained formally from
$|\beta\rangle_n\langle\bar{\beta}|=(-1)^{p_{\beta}}{
}_n\langle\bar{\beta}|\beta\rangle_n$, or exactly by introducing a
supertrace over the states in the $n$-th channel.

It is clear that this procedure to  express the total $R$-matrix
via its constituents can be generalized to any complicated
network. Below we are going to apply this technique to the bubble
and triangle diagrams.

\subsection{Action Formulation of the Networks}

{
{In this section we formulate the network problem in terms of the
integral kernels of the $R$-operators (formulas
\ref{2d-R-coherent},\ref{R-coherent})  in the space of the
Grassmanian functions and express an action principle for it.

 As in subsection $3.2$, we attach to each local
vertex with $m$-channels a fermionic vertex-function:
\begin{equation}
\label{n-ch} R_m(\{\bar{\psi}_i, \psi_j\})=e^{-\sum^m
r_{ij}\bar{\psi}_i \psi_j},
\end{equation}
 which is the kernel of the corresponding $R$-operator
(\ref{rn}), the $r_{ij}$ being the entries of the $r$matrix
describing the scattering process over that vertex. In order to
find the $r-$matrix($S-$matrix) corresponding to the entire
network, one must take the product of the $R$-operators
(\ref{2d-R-coherent},\ref{R-coherent}) corresponding to the
vertexes and integrate in the functional space over the internal
fermionic fields.

Suppose that we have a network with $n$ vertices, each of which
with $s_i,\;i= 1,\dots n$ internal links (which connect two
vertices of the network) and $q_i$ external links (which are
ending or starting in one vertex).  Any link in the diagrams has
its double with an opposite direction for the arrow). Let us
denote the Grassmann-variables connected with the links of the
$k$-th vertex as $\bar{\psi}^k_i,\;\psi^k_j$: by a convention the
variables $\bar{\psi}^k_i$ will be attached to the links with
outgoing arrows, while the variables $\psi^k_j$ attached to the
links with incoming arrows. Every internal link connects two
vertices possessing two pairs of the Grassmann-variables: in the
diagram below (Fig.7) they are $\bar{\psi}^k_i,\;\psi^k_i$ and
$\bar{\psi}^{k'}_{i'},\;\psi^{k'}_{i'}$.
{
%
%
%
\begin{figure}[h]
\centerline{\includegraphics[width=55mm,angle=0,clip]{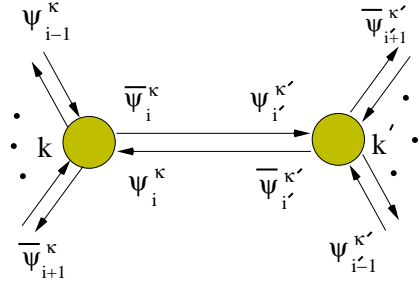}}
\caption{Attaching of the Grassmann-variables to the network. }
\end{figure}

 Note that  we are considering  general
networks, i.e.without any restriction on their structures,
 and, therefore, any
 two vertices can be connected by
an arbitrary number of links. Below we denote the set of
Fermi-fields on the internal links by $\mathcal{U}_{in}$, and the
set of Fermi-fields on the external links by $\mathcal{U}_{ex}$.

 Taking into account the integration rule
 for the product of integral kernels (\ref{product}),
 the vertex-function associated to the network will
be:
\begin{eqnarray}
&&{\cal R}(\{\bar{\psi}^p_i,{\psi}^{p'}_{i'}\}\in
{\mathcal{U}_{ex}} )=
e^{-\sum_{pp'}\sum_{i,i'\in{\mathcal{U}_{ex}}}\bar{\psi}^p_i {\bar S}^{p p'}_{ii'}\psi^{p'}_{i'}}=\nonumber\\
\label{int}&=&\frac{1}{N}
\int\prod_{\{i,i'\}\in{\mathcal{U}_{in}}}^{'}\prod_{k,k'}^n
d\bar{\psi}^k_i d \psi^{k'}_{i'}\prod_{k}^n R_k(\{\bar{\psi}^k_i, \psi^{k}_{i'}\})
\\
&=&
\frac{1}{N}\int\prod_{\{i,i'\}\in{\;\mathcal{U}_{in}}}^{'}\prod_{k,k'}^n\{d\bar{\psi}^k_i
d \psi^{k'}_{i'}\;e^{-\bar{\psi}^k_i \psi^{k'}_{i'}}\}
e^{-\sum_{i,j\in
 \mathcal{U}_{ex}\oplus\mathcal{U}_{in}}\sum_{k=1}^{n}r^{(k)}_{ij}\bar{\psi}^k_i
\psi^k_j}.\nonumber
\end{eqnarray}
Here $\mathcal{U}_{in}$ denotes also the set of internal links
between the vertices with $dim[\mathcal{U}_{in}]=\sum_j^n s_j=2s$,
where $s$ is the total number of the internal links in the
network. Similarly $\mathcal{U}_{ex}$ is the set of external links
and has dimension $q={{\sum_k^{n}q_k}}$, that is the number of the
external links of the network. The indices $k,i$ and $k',i'$ in
the measure of integration $e^{-\bar{\psi}^k_i \psi^{k'}_{i'}}$
correspond to the end points of the same link in the network
(hence the mark $'$ appearing on the symbols in the products of
(\ref{int}) and in the sums of (\ref{action}) below). The
normalization factor $N$ is present in order for the mega-network
$R$-matrix to appear in an exponential form (\ref{n-ch}) with an
identity pre-factor. It is defined by the contribution of the
closed loops to the functional integral (\ref{int}).

The argument of the exponential function in the integrant in
(\ref{int}):
\begin{equation}
\label{action} {\cal
A}(\{\bar{\psi}_i^k,\psi_{i'}^{k'}\}\in\mathcal{U}_{in} \oplus
\mathcal{U}_{ex} )= -\sum_{i,i'\in
\mathcal{U}_{in}}^{'}\sum_{k,k'=1}^{n}\bar{\psi}^k_i
\psi^{k'}_{i'} - \sum_{i,j\in
\mathcal{U}_{ex}\oplus\mathcal{U}_{in}}\sum_{k=1}^{n}r^{(k)}_{ij}\bar{\psi}^k_i
\psi^k_j
\end{equation}
can be regarded as an action, and the normalization factor $N$ as
a partition function for the corresponding network model with
fixed external (boundary) fields. This interpretation allows also
to calculate the $S$-matrix elements of the total network as a
Green function of the external Fermi-fields:
\begin{eqnarray}
  &(\mathcal{G})_{jj'}^{pp'} = ({\bar{S}}^{-1})_{j j'}^{p
  p'}&\\\nonumber
&=\frac{1}{N Det[\bar{S}]} \int
\psi^{p'}_{j'}\bar{\psi}^p_j
\prod_{\{i,i'\}\in{\mathcal{U}_{in}\oplus
\mathcal{U}_{ex}}}\prod_{k,k'}^n d\bar{\psi}^k_i d \psi^{k'}_{i'}
e^{-{\cal A}(\{\bar{\psi}_i^k,\psi_{i'}^{k'}\}\in\mathcal{U}_{in}
\oplus \mathcal{U}_{ex} )}.&
\end{eqnarray}

Now we will show that presented rules of calculation of the
$S$-matrix of the mega network reproduces the star-product
rules formulated in the articles \cite{KS1}. A general proof of equivalence
of two ways of calculations is available in a paper \cite{KSS}.
Here we will demonstrate their equivalence for so called bubble and
triangle diagrams. Moreover, we show that the same result can be obtained
via direct R-matrix products by following the rules presented in the
example (\ref{example}).

Let us separate  the summands in the exponent which contain
variables with internal indexes and represent the integral
(\ref{int}) in this way:
\begin{eqnarray}
\label{int-i}
\frac{1}{N}e^{-\sum_{i,j\in{\mathcal{U}_{ex}}}\sum_k^n
r^{(k)}_{ij}\bar{\psi}^k_i \psi^k_j}
\int\prod_{\{i,i'\}\in{\mathcal{U}_{in}}}\prod_{k,k'}^n
d\bar{\psi}^k_i d \psi^{k'}_{i'}\;\\\nonumber \cdot
e^{-\sum_{i,j\in{\mathcal{U}_{ex}}}
\sum_{i',j'\in{\mathcal{U}_{in}}}\sum_{k,k'}^n\left(\bar{\psi}^k_{i'}
{A_r}_{i'_{k} j'_{k'}} \psi^{k'}_{j'}+\bar{\psi}^k_{i} r^{(k)}_{i
i'}{\psi}^k_{i'}+\bar{\psi}^{k'}_{j'} r^{(k')}_{j'
j}\psi^{k'}_j\right)}.
\end{eqnarray}
where $A_{r}$ is a $2s\times 2s$ matrix, with the elements
${A_r}_{i_k j_{k'}}=\delta_{i_k j_{k'}}+\delta_{k k'} r^{(k)}_{i
j}$, where in the second summand only the elements relative to the
internal links are taken into account. This matrix, which contains
only unit elements on the diagonal, can be presented in a block
form after some rearrangement of the rows and columns , with $n$
diagonal blocks containing the parts of the $r^{(k)}$ matrices,
referred to the internal links of the network, and $2s$ unity
elements disposed out of the diagonal blocks in such a way that
each column and each row contain only one unity element. When
"tadpole"-vertices are present in the networks, elements of the
corresponding $r^{(k)}$ matrices also appear in the diagonal
together with the unity elements.

 After integration over internal $\psi's$ in the
Gaussian integral (\ref{int-i}), we arrive at:
\begin{eqnarray}\label{net}
 e^{-\sum_{p,p';i,i'\in{\mathcal{U}_{ex}}}\bar{\psi}^p_i {\bar S}^{p p'}_{ii'}\psi^{p'}_{i'}}=
\frac{\det{A_r}}{N}e^{-\sum_{i,j\in{\mathcal{U}_{ex}}}\left(\sum_k
r^{(k)}_{ij}\bar{\psi}^k_i \psi^k_j -\sum_{k',k''}
A^{\;\;k'k''}_{r\;ij}\bar{\psi}^{k'}_i {\psi}^{k''}_j \right)},
\end{eqnarray}
from which it follows that the normalization factor $N$ in
(\ref{int}) must be equal to $\det{A_r}$ and the mega-$S$-matrix
is:
\begin{equation}
{\bar S}^{k k'}_{ij}= r^{(k)}_{ij}\delta^{k k'} - A^{\;\;k
k'}_{r\;ij},
\end{equation}
where \bea \label{Ar} {A}^{\;\;k
k'}_{r\;ij}=\sum_{i',j'\in{\mathcal{U}_{in}}}^{2s}r^{(k)}_{i
i'}{(A_r)}^{-1}_{i'_{k}j'_{k'}}r^{(k')}_{j' j}
\end{eqnarray}
and $k, k'$ are relative only to vertices which possess external
legs.

Expression (\ref{net}) easily follows after using
Hubbard-Stratanovich transformation in (\ref{int-i}) and:
\begin{eqnarray}
&&\sum_{i',j'\in{\mathcal{U}_{in}}}^{2s}\left(\sum_{i\in{\mathcal{U}_{ex}}}^q\sum_k^n
\bar{\psi}^k_{i} r^{(k) }_{i
i'}\right){(A_r)}^{-1}_{i'_kj'_{k'}}\left(\sum_{j\in
{\mathcal{U}_{ex}}}^q \sum_{k'}^n r^{(k')}_{j' j}\psi^{k'}_j\right)=\nonumber\\
&=&\sum_{i,j\in{\mathcal{U}_{ex}}}^q \sum_{k,k'}^n A^{\;\;k
k'}_{r\;ij} \bar{\psi}^{k}_i {\psi}^{k'}_j. \eea

For the particular case $n=2$ let us write down a detailed
expression of the network vertex-function (or the kernel of the
$R$-operator corresponding to that network). Now the network
contains two vertices having respectively $s_1,\; q_1$ and
$s_2,\;q_2$ internal and external links. After dividing each of
the $r^{(k)}$ matrices into four blocks, following the definitions
in the work \cite{KS1}, the kernel function of  $k$-th vertex can
be represented as:
$$e^{-\sum_{i,j}^{s_k+q_k}r^{(k)}_{ij}\bar{\psi}^k_i
\psi^k_j}=e^{-\sum_{i,j}^2 {\mathrm{r}}^{(k)}_{ij}\bar{\Psi}^k_i
\Psi^{k}_j}$$
Here $\Psi^k_1=\{\psi^k_1,\cdots,\psi^k_{s_k}\}^{\tau}$ is the set
of variables connected with the internal links of the $k$-th
vertex, and
$\Psi^k_2=\{\psi^k_{s_k+1},\cdots,\psi^k_{s_k+q_k}\}^{\tau}$ the
set of variables connected with the external links.
Correspondingly:
$\bar{\Psi}^k_1=\{\bar{\psi}^k_{1},\cdots,\bar{\psi}^k_{s_k}\}$,
$\bar{\Psi}^k_2=\{\bar{\psi}^k_{s_k+1},\cdots,\bar{\psi}^k_{s_k+q_k}\}$.
It follows that:
 $r^{(k)}=\left(\begin{array}{cc}
{\mathrm{r}}^{(k)}_{11}& {\mathrm{r}}^{(k)}_{12}\\
{\mathrm{r}}^{(k)}_{21}&
{\mathrm{r}}^{(k)}_{22}\end{array}\right)$.

 Now the expression
(\ref{int}) takes a simple form (it is obvious that $s_1=s_2=s$):
\begin{eqnarray}\nonumber
& \displaystyle \int d\bar{\Psi}^1_1 d \Psi^{2}_{1}d\bar{\Psi}^2_1
d \Psi^{1}_{1} \exp\bigg\{-\bar{\Psi}^1_1
\Psi^{2}_{1}-\bar{\Psi}^2_1 \Psi^{1}_1 -\sum_{k=1}^{2}\sum_{i,j}^2
\bar{\Psi}^k_i{\mathrm{r}}^{(k)}_{ij}\Psi^{k}_j \bigg\}=&\\& N
\exp\Bigg\{\displaystyle-\sum_{k=1}^{2}
 {\mathrm{r}}^{(k)}_{22}\bar{\Psi}^k_2\Psi^{k}_2+\left(\bar{\Psi}^2_2
 {\mathrm{r}}^{(2)}_{21},\bar{\Psi}^1_2
 {\mathrm{r}}^{(1)}_{21}\right)A_r^{-1}
{\left(\!\!\begin{array}{c}
{\mathrm{r}}^{(1)}_{12}\Psi^{1}_2\\
{\mathrm{r}}^{(2)}_{12}\Psi^{2}_2\end{array}\!\!\!\right)}\Bigg\},&\label{int1}
\end{eqnarray}
where
\begin{eqnarray}
\label{AA}
&A_r=\left(\begin{array}{cc}\mathrm{I}&{\mathrm{r}}^{(1)}_{11}\\
{\mathrm{r}}^{(2)}_{11}&\mathrm{I}\end{array}\right),\qquad
N=\det{A_r}=\det{(\mathrm{I}-{\mathrm{r}}^{(1)}_{11}{\mathrm{r}}^{(2)}_{11})}.&
\end{eqnarray}
where $\mathrm{I}$ is the $s\times s$ unit matrix.

The quadratic function
 in (\ref{int1}) can be represent in a compact form as:
\begin{eqnarray}
e^{-\sum_{i,j}^{2}\mathrm{r}^{(12)}_{ij}\bar{\Psi^i_2}\Psi^j_2},
\end{eqnarray}
with the $(q_1+q_2)\times (q_1+q_2)$ matrix $\mathrm{r}^{(12)}$:
\begin{eqnarray}\label{s-p}
\mathrm{r}^{(12)}=\left(\begin{array}{cc}\mathrm{r}^{(1)}_{22}-
\mathrm{r}^{(1)}_{21}{[A_r^{-1}]}_{21}\mathrm{r}^{(1)}_{12}&-
\mathrm{r}^{(1)}_{21}{[A_r^{-1}]}_{22}\mathrm{r}^{(2)}_{12}\\&\\-
\mathrm{r}^{(2)}_{21}{[A_r^{-1}]}_{11}\mathrm{r}^{(1)}_{12}&\mathrm{r}^{(2)}_{22}
-\mathrm{r}^{(2)}_{21}{[A_r^{-1}]}_{12}\mathrm{r}^{(2)}_{12}
\end{array}\right),
\end{eqnarray}
and the inverse matrix  $A_r^{-1}$ having a following form
\begin{equation}
A_r^{-1}=\left(\begin{array}{cc}[A_r^{-1}]_{11}&{[A_r^{-1}]}_{12}\\&\\
{[A_r^{-1}]}_{21}&{[A_r^{-1}]}_{22}\end{array}\right)=
\left(\begin{array}{cc}\left(\mathrm{I}-{\mathrm{r}}^{(1)}_{11}{\mathrm{r}}^{(2)}_{11}\right)^{-1}
&-{\mathrm{r}}^{(1)}_{11}\left(\mathrm{I}-{\mathrm{r}}^{(2)}_{11}{\mathrm{r}}^{(1)}_{11}\right)^{-1}\\&\\
-{\mathrm{r}}^{(2)}_{11}\left(\mathrm{I}-{\mathrm{r}}^{(1)}_{11}{\mathrm{r}}^{(2)}_{11}\right)^{-1}&
\left(\mathrm{I}-{\mathrm{r}}^{(2)}_{11}{\mathrm{r}}^{(1)}_{11}\right)^{-1}\end{array}\right).
\end{equation}

 The results in (\ref{s-p}) reproduce for the considered situation
 the rules of the generalized
star-product, derived in \cite{KS1}. The matrix
$\mathrm{r}^{(12)}_{ij}$ describes the entire scattering process
over the $2-$vertex network. }}

 Now, let us note that the replacement of the local
$r^{(k)}$-matrices by their unitary counterparts is equivalent to
the unitary transformation of the mega $\bar{S}$-matrix. Here we
are giving the proof of that statement. We introduce formal
notation $\bar{\Psi}_{\mathcal{E}}S(r^{(k)})\Psi_{\mathcal{E}}
=\sum_{p,p';i,i'\in{\mathcal{U}_{ex}}}\bar{\psi}^p_i {\bar S}^{p
p'}_{ii'}\psi^{p'}_{i'}$,
 denoting by $\bar{\Psi}_{\mathcal{E}},\; {\Psi}_{\mathcal{E}}$
($\bar{\Psi}_{\mathcal{I}},\; {\Psi}_{\mathcal{I}}$) the fermionic
variables connected with the external (internal) links and by
${S}(r^{(k)})$ the global scattering matrix. We shall denote the
action in (\ref{action})by
$A(r^{(k)};\bar{\Psi}_{\mathcal{E}},{\Psi}_{\mathcal{E}},
\bar{\Psi}_{\mathcal{I}},{\Psi}_{\mathcal{I}})$.

Then, performing a simple transformation of the Grassman variables
\bea
N(r^{(k)})e^{-\bar{\Psi}_{\mathcal{E}}S(r^{(k)})\Psi_{\mathcal{E}}}=\int
D\bar{\Psi}_{\mathcal{I}} D\Psi_{\mathcal{I}}
e^{-A(r^{(k)};\bar{\Psi}_{\mathcal{E}},{\Psi}_{\mathcal{E}},
\bar{\Psi}_{\mathcal{I}},{\Psi}_{\mathcal{I}})}=\\\nonumber \int
D\bar{\chi}_{\mathcal{I}} D\chi_{\mathcal{I}}
e^{-A({r^{(k)}}^{\tau};\bar{\chi}_{\mathcal{I}},\chi_{\mathcal{I}}
,\bar{\chi}_{\mathcal{E}}\chi_{\mathcal{E}})}
=N({r^{(k)}}^{\tau})e^{-\bar{\chi}_{\mathcal{E}} S({r^{(k)}}^{\tau})\chi_{\mathcal{E}}},\\
{\textrm{where}} \quad {r^{(k)}}^{\tau}_{ij}={r^{(k)}}_{ji},\qquad
\bar{\chi}=-{\Psi},\quad \chi=\bar{\Psi},\eea
we are coming to the relations:
 $N({r^{(k)}}^{\tau})=N(r^{(k)})$
and $S({r^{(k)}}^{\tau})=S(r^{(k)})^{\tau}$.
The following relation takes also place:
$S({r^{(k)}}^{*})=S({r^{(k)}})^{*}$
where $*$ denotes complex conjugation.

If ${r^{(k)}}^{\dagger}={r^{(k)}}^{-1}$, then
$S({r^{(k)}}^{-1})=S({r^{(k)}}^{*\tau})=S({r^{(k)}})^{*\tau}$. As
$S(r^{(k)})$ is unitary, if $r^{(k)}$ is unitary matrix for each
vertex $k$ (see \cite{KSS}), then
\be S({r^{(k)}}^{-1})=S({r^{(k)}})^{-1}.\label{rs} \ee
and
$N({r^{(k)}}^{-1})=N(r^{(k)})^{*}$.

\section{Bubble Diagram}
Let us apply the technique developed above to the calculation of
the bubble network-diagram presented in Fig.7. According to the
prescription presented in the previous section the corresponding
two particle $R$-matrix is:



\begin{eqnarray}
 \label{babble-2}
{R^(2)}_{\alpha  \beta}^{\alpha_2\beta_2}
=\frac{1}{N}\sum(-1)^{p_{\gamma}}
{R^{(3)}}_{\alpha\;\;\beta\;\gamma }^{\alpha_1 \beta_1 \gamma_1}
{R^{(3)}}_{\alpha_1\;\beta_1\;\gamma_1 }^{ \alpha_2 \beta_2
\gamma},
\end{eqnarray}
where $N$ is defined from ${R^{(2)}}_{01}^{01}=1$, as it follows
from the expression of two particle R-matrix (\ref{XX}). This
gives
\begin{eqnarray}
 \label{N}
N= \sum(-1)^{p_{\gamma}} {R^{(3)}}_{0\;1\;\gamma }^{\alpha_1
\beta_1 \gamma_1} {R^{(3)}}_{\alpha_1\beta_1 \gamma_1 }^{ 0\; 1\;
\gamma}.
\end{eqnarray}

\begin{figure}[ht]
\centerline{\includegraphics[width=85mm,angle=0,clip]{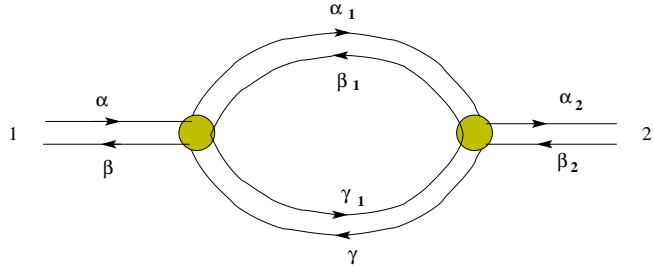}}
\caption{Bubble diagram constructed from two three channel vertices.}
\end{figure}

The calculations can be easily done with the Mathematica program.
In order not to overload the text with heavy expressions, we will
only present hereafter the result for scale invariant R-matrices
as defined in \cite{KS:1999}, \cite{BMS}, \cite{BBMS}. We begin by
stating the result which is particularly simple, namely:

Two scale invariant $R^{(3)}$-matrices with scaling parameters
$\varrho_1$ and $\varrho_2$ respectively, are defining, via the
expression (\ref{babble-2}), a scale invariant $R^{(2)}$ matrix
with scaling parameter $\varrho_1\varrho_2$.

More explicitly, according to papers \cite{KS:1999, BMS, BBMS} and
relation (\ref{r-matrix}), a scale invariant S-matrix defines the
following parameterisation for the R-matrix:
\begin{eqnarray}
\label{scale-1}
r_{11}&=&-\frac{\varrho}{a},\;\; r_{12}=\frac{1+\varrho}{a},\;\;
r_{13}=\frac{(1+\varrho)\varrho}{a},
\nonumber\\
r_{21}&=&\frac{(1+\varrho)\varrho}{a},\;\;
r_{22}=-\frac{\varrho}{a},\;\; r_{23}=\frac{1+\varrho}{a},
\nonumber\\
 r_{31}&=&\frac{1+\varrho}{a},\;\;r_{32}=\frac{(1+\varrho)\varrho}{a},\;\; r_{33}=-\frac{\varrho}{a},
\nonumber\\
a&=& 1+\varrho+\varrho^2
\end{eqnarray}

Substituting expressions (\ref{scale-1}) into (\ref{babble-2}) and
after renormalization of the product of the two $R^{(3)}$-matrices
by N one will obtain:
\begin{eqnarray}
 \label{scale-2}
r_{11}^{(2)}&=&\frac{2 \varrho_1 \varrho_2}{1+\varrho_1^2
\varrho_2^2},\qquad
r_{12}^{(2)}=\frac{1-\varrho_1^2 \varrho_2^2}{1+\varrho_1^2 \varrho_2^2}\nonumber\\
r_{21}^{(2)}&=&\frac{-1+\varrho_1^2
\varrho_2^2}{1+\varrho_1^2},\qquad r_{22}^{(2)}=\frac{2 \varrho_1
\varrho_2}{1+\varrho_1^2 \varrho_2^2},
\end{eqnarray}
which is precisely the parameterisation of the scale invariant two
particle S-matrix with scaling parameter $\varrho_1 \varrho_2$
defined in \cite{KS:1999, BMS, BBMS}!

 The bubble diagram we are considering is obviously the simplest
non-trivial $2$-vertex loop diagram. The $2 \times 2$ matrix
$r_{ij}$ can be constructed directly by use of both (\ref{s-p})
and (\ref{babble-2}) formulas.  Let us denote by $r_{ij}^{(k)}$
($k=1,\;2$) the two $3\times 3$ matrices, each corresponding to
one vertex. The numeration is taken so that the second channel of
the first vertex (internal links) corresponds to the first channel
of the second vertex, and the third channel of the first vertex is
connected with the third channel of the second vertex. The first
channel of the first vertex and the second channel of the second
vertex correspond  to the external links. The $r_{ij}$ matrix of
the final two channel process can be expressed by the following
expression \cite{KS1}:
\begin{eqnarray}
&r_{ij}=\left(\!\!\!\begin{array}{cc}r^{(1)}_{11}&0\\
0&r^{(2)}_{22}\end{array}\!\!\!\right)+
\left(\!\!\!\begin{array}{cccc}r^{(1)}_{12}&r^{(1)}_{13}&0&0\\
0&0&r^{(2)}_{21}&r^{(2)}_{23}\end{array}\!\!\!\right)\left[A_{12}^{-1}\right]
\left(\!\!\!\begin{array}{cc}r^{(2)}_{12}&0\\
r^{(2)}_{32}&0\\
0&r^{(1)}_{21}\\
0&r^{(1)}_{31}\end{array}\!\!\!\right),&
\end{eqnarray}
where
\begin{eqnarray}
&A_{12}=\left(\!\!\!\begin{array}{cccc}1&0&r^{(1)}_{22}&r^{(1)}_{23}\\
0&1&r^{(1)}_{32}&r^{(1)}_{33}\\
r^{(2)}_{11}&r^{(2)}_{13}&1&0\\
r^{(2)}_{31}&r^{(2)}_{33}&0&1\end{array}\!\!\!\right).&
\end{eqnarray}
The normalization factor $N$ equals to
\begin{eqnarray}
\label{NN2}
N&=&\mathrm{Det}[A_{12}]=\mathrm{Det}\left[\left(\!\!\!\begin{array}{cc}1&0\\
0&1\end{array}\!\!\!\right)-\left(\!\!\!\begin{array}{cc}r^{(1)}_{22}&r^{(1)}_{23}\\
r^{(1)}_{32}&r^{(1)}_{33}\end{array}\!\!\!\right)
\left(\!\!\!\begin{array}{cc}r^{(2)}_{11}&r^{(2)}_{13}\\
r^{(2)}_{31}&r^{(2)}_{33}\end{array}\!\!\!\right)\right]\qquad\nonumber\\
&=&1-r^{(1)}_{22}r^{(2)}_{11}-r^{(1)}_{33}r^{(2)}_{33}-
r^{(1)}_{32}r^{(2)}_{13}-r^{(1)}_{23}r^{(2)}_{31}\\\nonumber
&+&\left(r^{(1)}_{22}r^{(1)}_{33}-
r^{(1)}_{32}r^{(1)}_{23}\right)\left(r^{(2)}_{11}r^{(2)}_{33}-r^{(2)}_{13}
r^{(2)}_{31}\right).
\end{eqnarray}}
Note that the same expression (\ref{NN2}) for $N$ can be obtained
directly from formula (\ref{N}) by use of Mathematica program.

\section{Triangle Vertex}

In the most general case, the three particle R-matrix is
anisotropic and its elements depend on the channels 1,2,3.
In such a case,  it is necessary to specify, in the network
diagrams, the channels of all the three particle vertices and
follow which channels of neighboring vertices are connected. It is
clear that all $3^3=27$ types of connections in the triangle are
in general possible and this will be defined by the considered
problem itself. Here, as an example, we will present the case
corresponding of the disposition of vertices as depicted in Fig.8.
Namely, the second channel of the first vertex is connected with
the first channel of second vertex, the third channel of the
second vertex is connected with the second channel of the third
vertex and the first channel of third vertex is connected with the
third channel of the first vertex. Notice that, according to
Fig.5, the pairs of indexes ($\alpha , \beta $), ($\beta^{\prime},
\alpha^{\prime} $) and ($\gamma, \gamma^{\prime}$) at the first,
second and third position of ${R}$ are correspond to the first,
second and third channel respectively.

\begin{figure}[ht]
\centerline{\includegraphics[width=95mm,angle=0,clip]{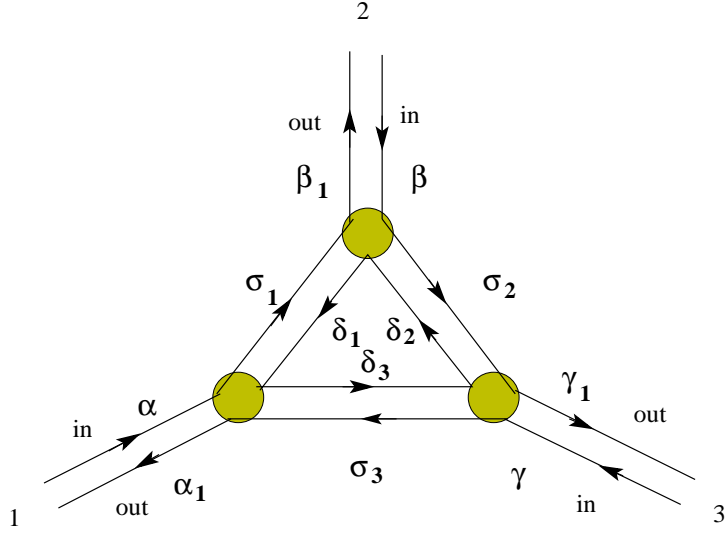}}
\caption{Triangle diagram constructed from three three-channel
R-matrices.}\label{f-triangle}
\end{figure}
Then, according to the prescription defined above, the
${R}$-matrix corresponding to triangle diagram reads as:
\begin{eqnarray}
\label{triangle-1} {R^{(\Delta)}}_{\beta_1 \alpha \alpha_1}^{\beta
\gamma_1 \gamma}= \frac{1}{N}\sum R_{\alpha \alpha_1
\sigma_3}^{\sigma_1 \delta_1 \delta_3} R_{\beta_1 \sigma_1
\delta_1} ^{\beta \sigma_2 \delta_2} R_{\sigma_2 \delta_2
\delta_3}^{\gamma_1 \gamma \sigma_3}(-1)^{p_{\sigma_3}}.
\end{eqnarray}
The normalization factor $N$ is defined by the condition
\begin{eqnarray}
 \label{N-2}
{R^{(\Delta)}}_{101}^{101}=1.
\end{eqnarray}


%
%

The matrix $r^{(123)}_{ij},\; i,j=1,2,3$ describing the scattering
over the triangle-graph in accordance with the formulas
(\ref{operator-R}) and (\ref{net}) is:
\begin{eqnarray}
&&r^{(123)}=\left(\begin{array}{ccc}r^{(1)}_{11}&0&0\\
0&r^{(2)}_{22}&0\\0&0&r^{(3)}_{33}
\end{array}\right) + \\
&+&\left(\begin{array}{cccccc}0&0&r^{(1)}_{12}&0&0&r^{(1)}_{13}\\
0&r^{(2)}_{21}&0&0&r^{(2)}_{23}&0\\r^{(3)}_{31}&0&0&r^{(3)}_{32}&0&0
\end{array}\right)A_{123}^{-1}\left(\begin{array}{ccc}r^{(1)}_{31}&0&0\\
r^{(1)}_{21}&0&0\\
 0&r^{(2)}_{12}&0\\
 0&r^{(2)}_{32}&0\\0&0&r^{(3)}_{23}\\
 0&0&r^{(3)}_{13}
\end{array}\right),\nonumber
\end{eqnarray}
where $r_{ij}^{(k)},\;\;k=1,2,3$ are the R-matrix parameters in
the three vertices of the triangle diagram on Fig.9. and
\begin{eqnarray}
A_{123}&=&\left(\begin{array}{cccccc}
1&0&r^{(1)}_{32}&0&0&r^{(1)}_{33}\\0&1&r^{(1)}_{22}&
0&0&r^{(1)}_{23}\\0&r^{(2)}_{11}&1&0&r^{(2)}_{13}&0\\
0&r^{(2)}_{31}&0&1&r^{(2)}_{33}&0\\r^{(3)}_{21}&0&0&r^{(3)}_{22}&1&0\\
r^{(3)}_{11}&0&0&r^{(3)}_{12}&0&1
\end{array}\right),
\end{eqnarray}
with
\begin{eqnarray}N=\det{A_{123}}.\qquad
\end{eqnarray}
Again, the calculations of ${{R}^{(\Delta)}}_{\beta_1 \alpha
\alpha_1}^{\beta \gamma_1 \gamma}$ can be done with a simple
Mathematica program. We will not present here the 20 big formulas,
but just show the normalization factor $N$:
\begin{eqnarray}
 \label{vacuum}\nonumber
N&=&\sum R_{0\; 1\; \sigma_3}^{\sigma_1 \delta_1 \delta_3} R_{1\;
\sigma_1 \delta_1} ^{1\; \sigma_2 \delta_2} R_{\sigma_2 \delta_2
\delta_3}^{0\; 1\; \sigma_3}(-1)^{p(\sigma_3)}\nonumber\\ &=&
\left(1-r^{(1)}_{22}r^{(2)}_{11}\right)\left(
1-r^{(1)}_{33}r^{(3)}_{11}\right)\left(1-r^{(2)}_{33}r^{(3)}_{22}\right)
\\\nonumber
&+&r^{(1)}_{23}
r^{(2)}_{31}r^{(3)}_{12}+r^{(1)}_{32}r^{(2)}_{13}r^{(3)}_{21}
-r^{(1)}_{33}r^{(2)}_{33}r^{(3)}_{12}r^{(3)}_{21}
\\
 \nonumber
&-&r^{(1)}_{22}r^{(2)}_{13}r^{(2)}_{32}r^{(3)}_{22}-r^{(1)}_{32}r^{(1)}_{23}r^{(2)}_{11}r^{(3)}_{11}+
 r^{(1)}_{22}r^{(1)}_{33}r^{(2)}_{11}r^{(2)}_{33}r^{(3)}_{11}r^{(3)}_{22}
\\\nonumber
&-&\left(r^{(1)}_{22}r^{(1)}_{33}-r^{(1)}_{23}r^{(1)}_{32}\right)
 \left(r^{(2)}_{11}r^{(2)}_{33}-r^{(2)}_{13}r^{(1)}_{31}\right)
 \left(r^{(3)}_{11}r^{(3)}_{22}-r^{(3)}_{12}r^{(3)}_{21}\right).
\end{eqnarray}

\section{Conclusions}

We have presented a new approach to the calculation of complicated
mega-S-matrices of the arbitrary network based on two ways:
\begin{itemize}
\item Introducing R-matrices
which are defined by the S-matrices of the local scatterings and taking the
product of such R-matrices;
\item Formulating an action for the
corresponding network by use of fermionic fields.
\end{itemize}
In the examples of bubble and triangle diagrams we show the
equivalence of this technique of computations with the generalized star-product procedure.
The general proof  valid in the case of an arbitrary graph
 was given in \cite{KSS}.

The action formulation has several advantages. First of all, it
allows one to investigate  network models with large  number ($N\rightarrow
\infty$) and  random disposition of vertices and formulate
corresponding quantum field theory, which will boost further
investigations. In this language one can take also sum
over random dispositions of scatterers and analyze the problem of
Anderson localization. Secondly, the approach makes the
calculation of mega-$S$-matrix elements of the regular,i.e. having
some translational symmetry,  networks very simple. Then the
problem simply reduces to Fourier analysis of the action on a
particular lattice defined by the network. This statement was
demonstrated in Section 2 on a model of a chain of equidistant
impurities. The third, and presumably most important advantage of
the action formulation, is a possibility to generalize networks
and consider models with multiparticle interactions.

\section{Acknowledgments}
We appreciate discussions with  M.Karowski,  M.Mintchev and
R.Schrader. A.S. acknowledge LAPTH and University of Savoie for
hospitality, where part of this work was done. P.S. is indebted to
L.Alvarez-Gaum\'e and the CERN Theory Division for its stimulating
and warm atmosphere.

\end{document}